# The photoinduced hidden metallic phase of monoclinic $VO_2$ driven by local nucleation via a self-amplification process

Feng-Wu Guo[1,2]†, Wen-Hao Liu[1]†, Zhi Wang[1], Shu-Shen Li[1,2], Lin-Wang Wang[1]*, and Jun-Wei Luo[1,2]*

[1]State Key Laboratory of Superlattices and Microstructures, Institute of Semiconductors, Chinese Academy of Sciences, Beijing 100083, China

[2]Center of Materials Science and Optoelectronics Engineering, University of Chinese Academy of Sciences, Beijing 100049, China

†These authors contributed equally to this work.
*Email: lwwang@semi.ac.cn; jwluo@semi.ac.cn

**Abstract:**

The insulator-to-metal transition (IMT) in vanadium dioxide ($VO_2$) has garnered extensive attention for its potential applications in ultrafast switches, neuronal network architectures, and storage technologies. However, a significant controversy persists regarding the formation of the IMT, specifically concerning whether a complete structural phase transition from monoclinic (M1) to rutile (R) phase is necessary. Here we employ the real-time time-dependent density functional theory (rt-TDDFT) to track the dynamic evolution of atomic and electronic structures in photoexcited $VO_2$, revealing the emergence of a long-lived monoclinic metal phase (MM) under low electronic excitation. The emergence of the metal phase in the monoclinic structure originates from the dissociation of the local V-V dimer, driven by the self-trapped and self-amplified dynamics of photoexcited holes, rather than by a pure electron-electron correction. On the other hand, the M1-to-R phase transition does appear at higher electronic excitation. Our findings validate the existence of MM phase and provide a comprehensive picture of the IMT in photoexcited $VO_2$.

The intricate interplay of charge, spin, lattice, and orbital degrees of freedom leads to a variety of emergent phenomena in strongly correlated materials and may also create a barrier to learning their underlying mechanisms [1-5]. For instance, the insulator-to-metal transition (IMT) behaviors accompanied by a rapid, reversible structural phase transition from the monoclinic (M1) insulating phase to the rutile (R) metallic phase in vanadium dioxide ($VO_2$) [6-10] have gained great attention in the past three decades due to its potential applications in ultrafast switches, neuronal network architectures, and storage technologies [11-14]. At low temperatures, $VO_2$ is stabilized in the M1 phase and is characterized by the Peierls V-V dimerization followed by titling to form zigzag chains (Figs. 1a, f and Supplementary Fig. 1a). As the temperature rises above 341 K, it transforms to the R phase [15,16] due to the dissociation of the V-V dimers (Figs. 1d, g and Supplementary Fig. 1b). Photoexcitation using ultrafast laser pulses provides a powerful approach to investigate and manipulate the phase transition of $VO_2$ with a timescale in the limit of atomic motion [17-24]. It is widely acknowledged that surpassing a certain threshold of laser fluence induces the IMT in $VO_2$, followed by a gradual lattice reorganization from the M1 phase to the R phase [4,17,25-29], which is attributed to electron-phonon coupling [30,31].

However, recent experiments have unveiled that a weaker laser fluence than the threshold required for M1-to-R phase transition can also induce the IMT, resulting from the formation of a metastable monoclinic metal (MM) phase [4,11,29,32-35]. Specifically, Morrison et al. [4] demonstrated the instantaneous emergence of the MM phase within a picosecond lifetime by employing a combination of ultrafast electron diffraction (UED) and time-resolved infrared (IR) transmittance. Sood et al. [11] further indicated that the electric drive can also induce such a long-lived MM phase in addition to photoexcitation. The emergence of this MM phase is commonly attributed to the suppression of the electron-electron correlation at low photoexcitation [4,11,29,32,33], reflecting the perception that conventional photoinduced IMT is driven by Mott-type transition. On the other hand, Xu et al. [36] refuted the existence of the MM phase, arguing that there is only one threshold for photoinduced phase transition and thus no separate threshold for this MM phase. This refutation was also supported by the absence of the MM phase in structural measurements of the femtosecond total X-ray scattering [26] and optical spectra utilizing the X-ray absorption spectroscopy, broad-band optical detection [28], and time- and spectrally resolved coherent X-ray imaging [21]. This discrepancy in the observations of the MM phase is vaguely argued to be the differences of the $VO_2$ samples (single-crystal [26,36,37] or polycrystalline [4,11,29]), and the overlooked differences in heat dissipation between different instruments [28]. Hence, it is urgent to carry out a theoretical investigation to examine the existence of the MM phase and reveal the underlying microscopic mechanism.

In this work, aimed at resolving the controversy surrounding the existence of the relatively long-lived MM phase, we employ the rt-TDDFT simulations to investigate the IMT dynamic processes. Under strong photoexcitation, the photoexcited holes induce significant atomic driving forces on each V-V dimer, breaking all V-V dimers and resulting in the formation of the metallic R phase (Fig. 1b), as discussed in our previous works [9]. Conversely, under weak photoexcitation, the photogenerated holes are insufficient to generate the necessary atomic driving force to dissociate all V-V dimers simultaneously (Fig. 1c). However, we observe the dissociation of a specific pair of V-V dimers (labeled as V3-V4) within the supercell (Fig. 1e). This local dissociation of the V-V dimer is attributed to the transfer of photoexcited holes to the vicinity of two V atoms, amplifying the atomic driving force acting on them and causing an elongation of their bond length. Importantly, this local structural phase transition will not alter the monoclinic-phase characteristics of the overall crystal structure under the XRD measurements [38,39], but, results in the localization of electronic wave function (Fig. 1e), corresponding to an isolated band crossing the Fermi level (Fig. 1h and Supplementary Fig. 1c) and giving rise to the formation of MM phase within a picosecond lifetime. It is crucial to emphasize that the underlying MM phase is associated with atomic structure changes rather than purely electronic effects as previously believed [4,11,29,32,33]. This clarification sheds light on the nature of the MM phase characterized by local structural alterations in $VO_2$, thereby enhancing our comprehension of the photoinduced IMT and laying the groundwork for future exploration in this field.

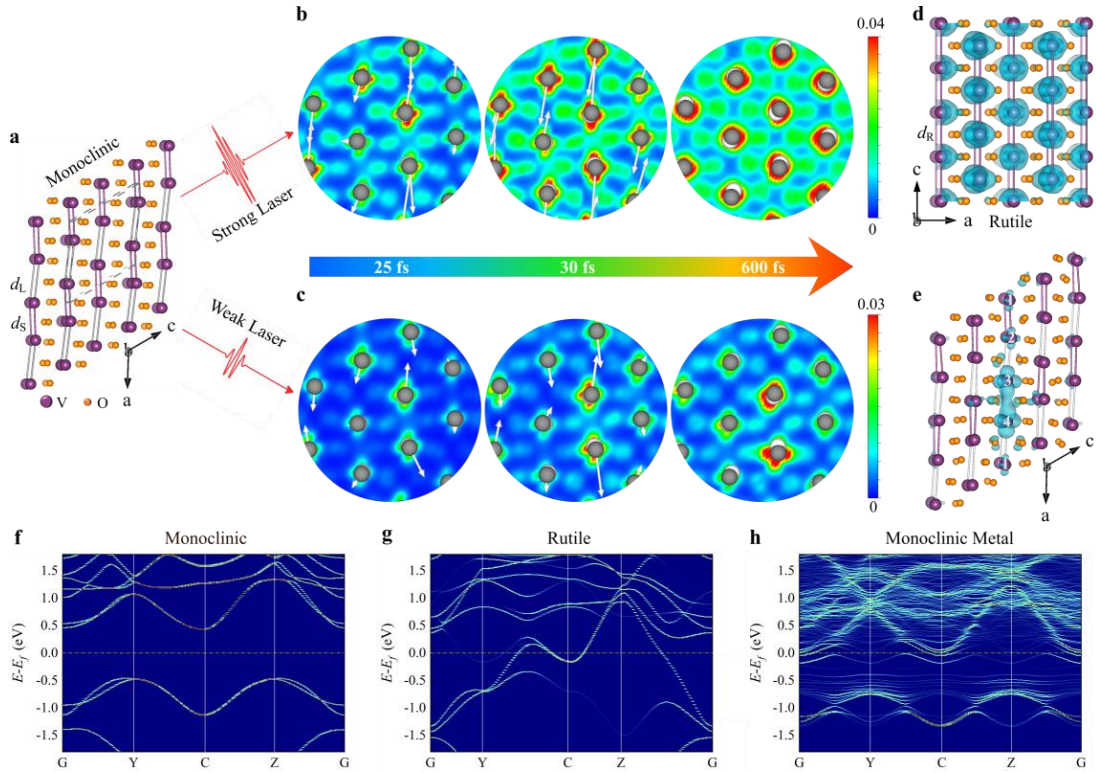

**Fig. 1. Fluence-dependence of photoinduced phase transitions in VO$_2$. a,** The atomic structure of the 2 × 2 × 2 supercell for the insulating monoclinic (M1) phase. Here, V and O atoms are labeled by purple and orange balls, respectively, and the dashed gray parallelepiped represents the unit cell. The Peierls lattice distortion along the a-axis causes V-V dimerization with the bonds of the V-V dimers (with a bond length $d_S$) and the bonds of long counterparts (with a bond length $d_L$) represented by purple lines and gray lines, respectively. **b,c,** The time evolution of the real-space distribution of photoexcited holes on the (010) plane for M1-phase VO$_2$ following (b) a strong photoexcitation and (c) a weak photoexcitation, respectively. For comparison, the initial positions of the V atoms before photoexcitation are indicated by white balls, and the photoinduced atomic driving forces on V atoms are marked by white arrows with their length indicating the relative force strength. **d,** The corresponding 2 × 2 × 4 supercell of the metallic rutile (R) phase, in which all V-V bonds (purple lines) along the a-axis are equally distributed at $d_R$. The blue shading indicates a 3D isosurface of the partial charge density of the bands around the Fermi level as marked in (g). **e,** The 2 × 2 × 2 supercell structure of the metal monoclinic (MM) phase with a 3D isosurface of the partial charge density of an isolated band around the Fermi level marked in (h). The V atoms in a chain with larger displacements are labeled by Arabic numerals 1-4. **f-h,** The first-principles calculations predicted unfolded band structures of the M1 phase (f), the R phase (g), and the MM phase (h).

## Results

**Photoinduced M1-to-R phase transition.**

Figure 2 shows the temporal evolution of band gap, V-V bond lengths, and simulated X-ray Diffraction (XRD) of the M1-phase VO$_2$ at 300 K following the photoexcitation at fluences corresponding to excitations of 0.79%, 0.97%, and 1.25% of valence electrons to conduction bands. These fluences are near the threshold for the conventional M1-to-R structural phase transition in VO$_2$ [9]. One can see from Fig. 2f that, at strong photoexcitation of 1.25%, bond lengths of all V-V dimers get longer and that of their long bond counterparts get shorter immediately after the photoexcitation. At around 120 fs, these bond lengths become equally distributed at 2.74 Å, which corresponds to the V-V bond length in the R phase. After that, all bonds still vary slightly around 2.74 Å in damped oscillations. From the evolution of V-V bonds, we can identify that the VO$_2$ system undergoes a photoinduced transition from the M1 phase to the R phase. To further confirm it, we examine the XRD pattern in which the vanishing of the $30\bar{2}$ Bragg peak and the slight intensity increase of the 200 and 220 peaks signify the transition from the M1 phase to the R phase [4,26,29,36]. Figure 2i shows that the evolution of the 200, 220, and $30\bar{2}$ Bragg peaks in our simulated XRD pattern based on the dynamic evolution of atom positions (Supplementary Figs. 3 and 4) is consistent with the experimental measurements for photoinduced M1-to-R phase transition[36]. Specifically, our theoretically predicted intensity of the $30\bar{2}$ peak disappears completely at ~150 fs and then becomes steady without any oscillation (Fig. 2i), indicating a complete phase transition to the R phase [26,36]. Note that the slight differences in intensity and time scale between our simulation and experiments may arise from differences in photoexcitation strength and sample imperfection [37,40]. From the band structure of VO$_2$ in the R phase (Fig. 1g), one can see that multiple energy bands intersect the Fermi level and close the band gap. Figure 2c shows that right after the photoexcitation the bandgap drops sharply and approaches zero after 50 fs, along with the crystal structure transition toward the R phase. The resulting charge modulation will further stabilize the VO$_2$ in the R phase, giving rise to no coherent phonon oscillation between the two phases involved in the transition, which is frequently observed in the experimental literature [26,36].

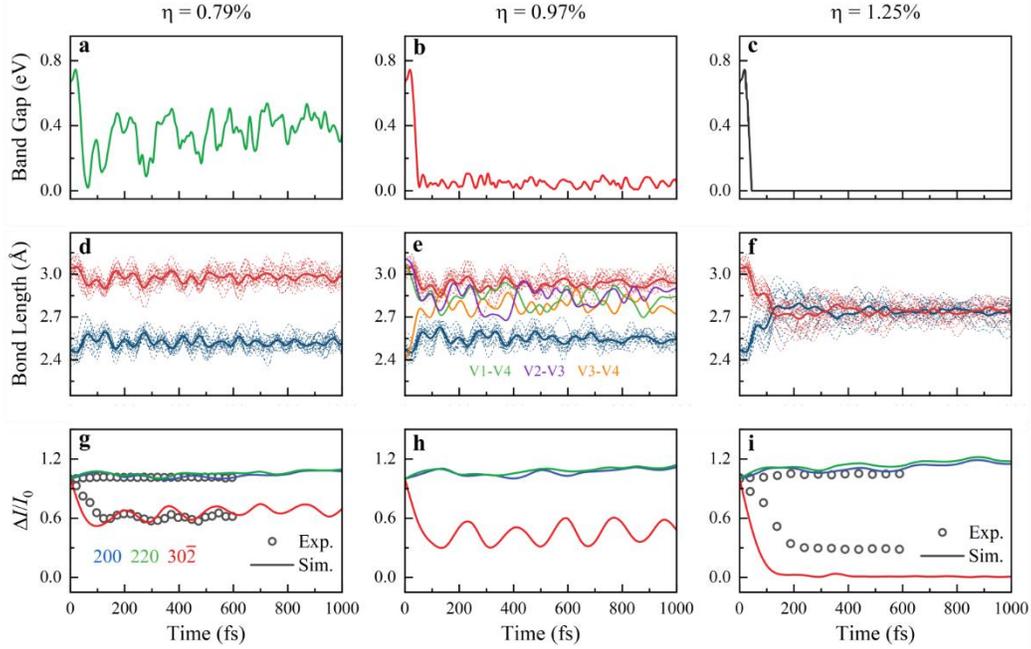

**Fig. 2 Temporal evolution of band gap and atomic structure of the M1-phase VO$_2$ following photoexcitation. a-c,** The dynamic evolution of the band gap at photoexcitation of η = 0.79% (a), η = 0.97% (b), η = and 1.25% (c). η represents the proportion of valence electrons photoexcited to the conduction bands. **d-f,** the corresponding temporal dynamics of the V-V bond lengths. The red and blue dashed lines correspond to the long and short bond lengths, respectively. The average bond lengths of long and short bonds are represented by red and blue solid lines, respectively. The green, purple, and orange solid lines in (e) highlight the bonds of a local chain marked by Arabic numerals 1-4 in Fig. 1(e). **g-i,** The corresponding time evolution of the simulated 200 (blue lines), 220 (green lines), 30$\bar{2}$ (red lines) Bragg peaks compared with available experimental data (black circles) [36].

**Photoinduced MM phase.**

As the photoexcitation strength is reduced to 0.97% excitation, Figure 2b illustrates that the bandgap rapidly diminishes to zero, akin to the behavior observed at 1.25% photoexcitation, indicating the presence of a long-lived metallic state. However, in this case, the bandgap then oscillates around zero, albeit with distinctions from the case at 1.25% photoexcitation. Notably, the corresponding unfolded band structure (Fig. 1h) reveals an isolated band near the Fermi level when projecting the electronic states of the VO$_2$ system at 500 fs. Figure 1e shows that this isolated band corresponds to the electronic states highly localized on one of the V-V dimers (associated V atoms denoted as V3 and V4, respectively). Figure 2e shows that the bond length of this V-V dimer (V3-V4) and that of the corresponding long counterparts (V2-V3 and V4-V1) become equal at 110 fs and then rebound slightly with an oscillation thereafter but never recover back to their original bond lengths. Meanwhile, the remaining V-V dimers have a very small change in bond length and maintain the characteristic features of the M1 phase (Figs. 1e and 2e). The reduced intensity of the simulated $30\bar{2}$ peak drops by about 55% at 150 fs and then slightly recovers (Fig. 2h), exhibiting an oscillation of 6 THz consistent with the coherent phonon observed experimentally [17,36,40-42]. Therefore, we demonstrate that the formation of the metastable metallic (MM) phase under 0.97% photoexcitation is driven by a local structural transition rather than driven by electron-electron correlation without structure change as previously believed [4,11,29,32,33]. This finding of photoinduced metallic metastable phase arising from local atomic nucleation has been postulated very recently to explain the experimental data for the photo-susceptibility of $VO_2$ [43] and time-resolved harmonic spectroscopy in $NbO_2$ [44].

The existence of the MM phase has been experimentally confirmed through the analysis of changes in the transmission spectrum [4]. To deepen understanding, we have modified the Random Phase Approximation (RPA) method [45] to calculate the absorption spectra of photoexcited states (Supplementary Note 3). We then extract the values of the absorption spectra at the photon energy $\hbar\omega = 0.25$ eV [4] to generate the transmission spectrum (Supplementary Fig. 8). At photoexcitation of 0.97%, the IR transmission reaches a minimum and almost remains at that value due to the long-lived existence of metallic properties with an isolated band near the Fermi level, which also aligns well with experimental results [4]. Besides, one hallmark of the MM phase observed in previous experiments on polycrystalline samples is the rapid decrease in the $30\bar{2}$ peak followed by a slow increase in the 200 peak [4,11,29]. However, this delay effect has been subjected to debate in other experiments, with the contention that it is only observed in polycrystalline samples [4,11,29] and cannot be reliably detected in single crystal samples [26,36,37]. Consequently, it is not rigorous to discern the occurrence of a metastable phase solely based on the delay phenomenon observed in the 200 and $30\bar{2}$ Bragg peaks. We also propose that future experiments can utilize time-resolved angle-resolved photoemission spectroscopy (tr-ARPES) [19,46,47] to discern the isolated band near the Femi level, thereby validating the presence of the MM phase.

Further reducing the photoexcitation to 0.79%, Figure 2a shows that the magnitude of the bandgap falls sharply, approaching almost zero at 60 fs after photoexcitation, but then quickly increases to 0.3 eV followed by a damped oscillation within the range from 0.1 to 0.5 eV thereafter.

Along with this behavior of bandgap, the bond lengths of the V-V dimers get longer, and their long bond counterparts get shorter, tending to their bond length of 2.74 Å in the R phase within the initial 100 fs. Before getting equal, they rebound partially with long-lived oscillations after 100 fs (Fig. 2d). Particularly, different from the above two cases of strong photoexcitation, there is no V-V bond approaching the bond length of 2.74 Å in the R phase. Figure 2g shows that our simulated XDR pattern is in excellent agreement with experimental measurements [36]. One can see from Fig. 2g that the $30\bar{2}$ peak in the simulated XDR pattern (see Supplementary Note 2 for details) exhibits a drop of approximately 35% in intensity with an oscillation corresponding to a 6.0 THz coherent phonon mode [17,36,40-42], indicating that the crystal structure is still maintained in the M1 phase [36]. Consequently, at 0.79% photoexcitation a transient metal phase is emerging but is, however, last only for a few tens of femtoseconds.

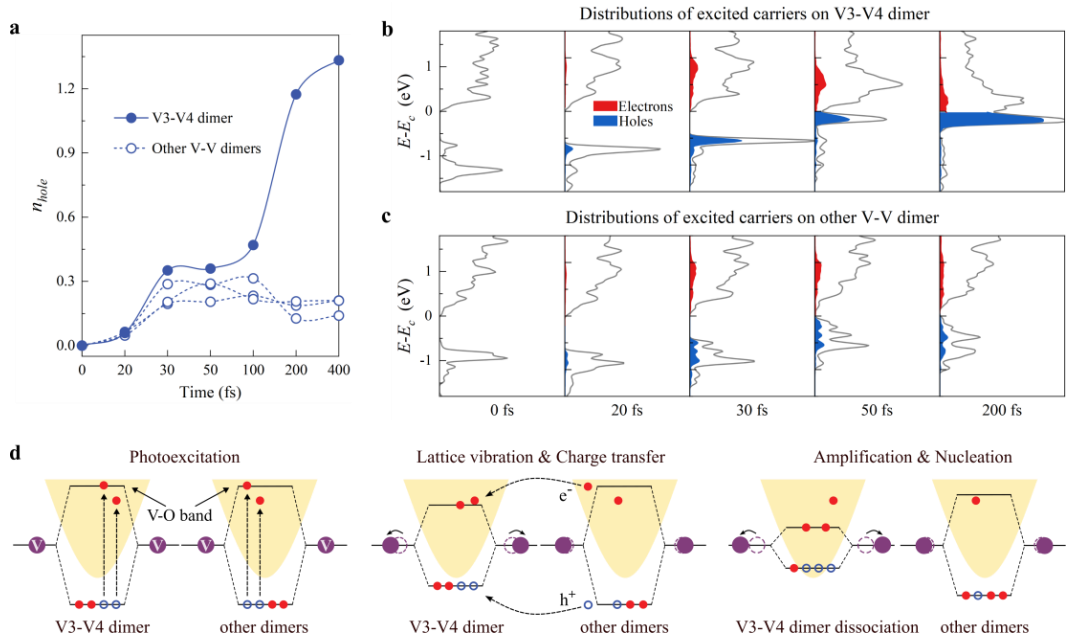

**Fig. 3. Distributions of photoexcited carriers on V-V dimers. a,** The evolution of photoexcited holes on V-V dimers at electronic excitations of 0.97% within -1.6-0 eV below the Fermi Level. The V3-V4 dimer is represented in a solid blue line, and another V-V dimer is illustrated in dashed blue lines. **b, c,** The projected density of states (PDOS) of V3-V4 dimer (b) and other V-V dimers (c) from TDDFT simulations at different times. Red (blue) shaded areas represent the distributions of photoexcited electrons (holes). **d,** The schematic diagram of the self-trapping and self-amplified dynamics.

## Discussion

**Origin of photoinduced MM phase.**

We have demonstrated that the experimentally observed long-lived metallic metastable phase [4,11,29] indeed exists and occurs within a narrow photoexcitation range from 0.79% to 1.25%. To reveal the mechanism underlying the local nucleation-driven metallic metastable phase near just below the threshold photoexcitation for M1-to-R phase transition, we recall the recently uncovered photoexcitation-dependent atomic driving forces for such structural phase transition [9], which unifies the ordered coherent [27] and disorder [26] dynamics observed experimentally in photoexcited $VO_2$. It was found [9] that the photoexcited holes generate a force on each V-V dimer to drive their collective coherent motion, which is in competition with the thermal-induced random vibrations. Specifically, the M1 phase $VO_2$ under photoexcitation with photon energy (1.55 eV) above bandgap, $n$ valence electrons from the $d_{\parallel}$ bonding states of V-V dimers are vertically excited to the $d_{\parallel}^*$ antibonding states of V-V dimers and $\pi^*$ antibonding states of V-O bonds in the conduction bands (Fig. 3d and Supplementary Fig. 13). Such charge modulation between bonding states and antibonding states increases the free energy of the $VO_2$ system in an amount proportional to $nV$, where the overlap parameter $V$ is inversely proportional to the square of the bond length of the V-V dimers [9]. An interatomic force on each V-V dimer is thus generated in order to elongate the bond length of V-V dimers by lowering the free energy of the photo-excited system. A strong laser fluence excites more valence electrons $n$ and thus generates stronger atomic forces. Meanwhile, the V atoms are in random motions around the equilibrium lattice sites driven by thermally excited phonons. At a very strong photoexcitation strength, the photogenerated atomic driving forces are much larger than the perturbation atomic forces due to thermal phonons and drive the atoms toward the M1-to-R structural phase transition direction in a deterministic way, and the thermal phonons are only a small perturbation. As the photoexcitation strength is lowered approaching the threshold (e.g., 1.25% photoexcitation), the photogenerated atomic forces become comparable with the atomic forces associated with thermal phonon vibrations, making the M1-to-R structural phase transition process disordered due to V atoms being pushing away from the transition path by thermal phonon vibrations, as shown in Fig. 1b.

If the photoexcitation strength is reduced slightly below the threshold where the photogenerated atomic forces are not strong enough to drive the M1-to-R structural phase transition, a different dynamic emerges. To track the evolution of the photoexcited holes on each individual V-V dimer, we define $p_h(t)$ as the integration of electronic states located on each V-V dimer weighted by time-dependent unoccupation number $\rho_{V-V}(E,t)$ within a range from -1.6 eV to the Fermi level (located at 0 eV), as shown in Figs. 3b, c: $p_h(t) = \int_{-1.6eV}^{0} \rho_{V-V}(E,t)dE$. Figure 3a shows $p_h(t)$ for different V-V dimers. One can see that all $p_h(t)$ increase at a nearly equal rate at the initial 30 fs arising from the photoexcitation by a laser pulse. After that, they tend to have a relatively steady value for V-V dimers except for the V3-V4 dimer, which gets larger even after the photoexcitation (Fig. 3b). It implies that the enhanced part of photoexcited holes on the V3-V4 dimer relative to the remaining V-V dimers should arise from a transfer from its nearby area (Fig. 1c and Supplementary Fig. 12a). It has been revealed that the electron-phonon coupling-induced self-amplification process creates homogeneous atomic nucleation in photoinduced nonthermal melting in Si [48]. Figure 3d schematically shows that the thermal phonon vibrations could induce a local maximum, say at the V3-V4 dimer, in the energy landscape of the valence band edge according to the deformation potential theory originally developed by Bardeen and Shockley [49]. This local maximum renders the V3-V4 dimer as a hole-trapping center to capture more photoexcited holes from its vicinity, as shown in Fig. 3b and diagramed in Fig. 3d, which further enhances the atomic force on the V3-V4 dimer alone, as shown in Fig. 1c. Larger atomic force drives the bond length of the V3-V4 dimer longer than the remaining V-V dimers to have a deeper potential well for holes, as shown in Fig. 3a and schematically diagramed in Fig. 3d. The deeper well can capture more holes and so on, forming a self-amplification process induced by electron-phonon coupling. Such a self-amplification process successively upraises the bonding state band of the V3-V4 dimer and increases the number of localized holes on the V3-V4 dimer, as shown in Fig. 3a, yielding a sufficient strong atomic force to dissociate the V3-V4 dimer but sustaining the remaining V-V dimers. Such self-amplified local distortion generates atomic nucleation, which is responsible for the emergence of the long-lived MM phase as observed in the simulation case of 0.97% photoexcitation (Fig. 1c and Fig. 2b) and experimentally measurements [4,11,29]. Note that this photoinduced local atomic nucleation is a bit like the formation of polarons, which has been identified as the main factor that causes the metallic states as reported in other materials such as in $Nd_2CuO_4$ [50] and $BaBiO_3$ [51,52]. It will be extremely helpful to use the experimental technique used in photoexcited polarons observation [53-56], such as the transient extreme-ultraviolet spectroscopy on an ultrafast timescale, to probe the MM phase in $VO_2$.

Because electron-phonon coupling plays a central role in such self-amplification processes, such photoinduced atomic nucleation is naturally dependent on both photoexcited holes and thermal phonons. Supplementary Figure 12b shows that if the initial temperature of $VO_2$ lowers to 100 K to suppress the thermal phonon vibrations, the photoinduced local atomic nucleation does not appear, and the photoexcited holes and atomic driving force are almost evenly distributed at all V-V dimers. On the other hand, if we lower the photoexcitation to 0.79% excitation, a certain V-V dimer can only capture more photoexcited holes by tens of femtoseconds but is insufficient to dissociate corresponding V-V dimer (Supplementary Fig. 14). Figure 2a shows that this transient capture of photoexcited holes results in the instantaneous collapse of bandgap (emergence of a metallic state) followed by a rapid recovery process (see also Supplementary Fig. 14).

**Role the electron-electron correlation and electron-phonon coupling.**

Although we have demonstrated the IMT in the M1 structure being driven by the structural change in the atomistic structure, it is also worth examining the photoinduced electron correlation effect on the bandgap since the IMT in $VO_2$ is also usually regarded as a Mott-type transition, where electron correlations lead to a collapse of the band gap [4,11,29,32,33]. This is particularly interesting considering the experimental approaches are difficult to access the electron correlation effect alone due to the ever existence of electron-phonon coupling in the system. Here, we investigate this question theoretically using rt-TDDFT simulations by freezing the atomic positions, thus turning off the electron-phonon coupling. Figure 4 shows that the bandgap of M1 phase $VO_2$ gets smaller linearly even without structural change by increasing the photoexcitation strength, indicating the effect of electron correlation effect on the bandgap. The bandgap collapses completely at a strong photoexcitation of about 3.75% valence electrons pumped into the conduction bands, indicating that the photoexcitation threshold for the transition to the metal phase driven by the electron correlation is much higher than that driven by the local nucleation (Fig. 4a and Supplementary Note 1, Figs. 2c, d) and thus latter is preferred in the competition between these two mechanisms. This predicted threshold for the photoinduced Mott-type transition is even higher than that for the conventional M1-to-R phase transition in $VO_2$ [4,11,29,34].

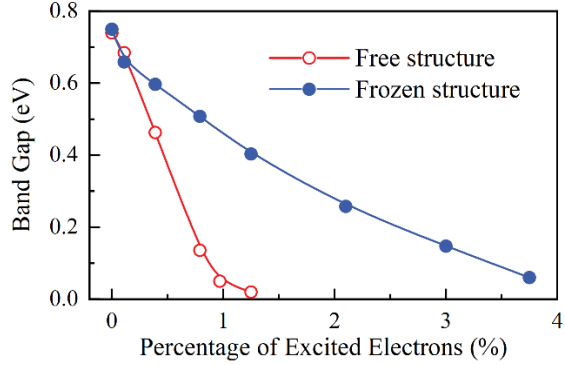

**Fig. 4.** The band gap of the VO$_2$ at 50 fs following the photoexcitation as a function of the percentage of excited valence electrons in comparison with that narrowed by the electron correlation effect at the same photoexcitation but frozen the atomic structure to the M1 phase.

In summary, our rt-TDDFT simulation confirms the existence of experimentally observed long-lived MM phase in VO$_2$ [4,11,29,43] under photoexcitation slightly weaker than the threshold for the M1-to-R structural transition. Specifically, under a strong photoexcitation above the threshold for the M1-to-R structural transition, the photoexcited holes occupied the bonding state band of the V-V dimers to generate an atomic driving force on each V-V dimer to elongate its bond until the dimer is completely dissociated [9], and the dissociation of all V-V dimers results in the transition to the metallic R phase (Fig. 1b). Conversely, under weak photoexcitation, the photogenerated holes are insufficient to generate the necessary atomic driving force to dissociate all V-V dimers simultaneously (Fig. 1c). However, we observe that one of V-V dimers in the supercell experience slight stretching due to the influence of thermal phonons (Fig. 3d). This subtle displacement induces the transfer of photoexcited holes to the vicinity of two V atoms, amplifying the atomic driving force acting on them to further stretch their bond length. The process increases the local well depth and can trap more holes from the vicinity, creating a self-amplified and self-trapped process. This interplay between photoexcited holes, atomic driving forces, and V-V dimer displacement initiates a self-amplification in the dynamical process, eventually leading to the complete dissociation of a pair of V-V dimer leaving the remaining dimers intact. The localized effect observed in the MM phase contributes to its relatively long-lived lifetime, which justifies its appropriate assignment as an IMT phase based on electronic structure analysis. In contrast to prior literature assuming no structural changes, we demonstrate that the emergence of a metallic state in the M1 phase arises from the local atomic nucleation, which is consistent with very recent experimental evidence observed in VO$_2$ [43] and NbO$_2$ [44]. From the view of the photogenerated atomic driving force, we develop a unified understanding of both photoinduced M1-to-R phase transition at strong photoexcitation and M1-to-MM phase transition at low photoexcitation.

## Methods

All calculations were performed using the ab initio package PWmat [57], incorporating the local density approximation functional (LDA) + U exchange-correlation [58]. Our simulation model comprised a 96-atom supercell, employing a plane-wave basis with a cutoff energy of 45 Ry. To ensure accurate Brillouin-zone sampling, a 2×2×2 Monkhorst-Pack k-point mesh was utilized. A Hubbard U value of 3.4 eV for V 3d states was employed, yielding a bandgap of approximately 0.7 eV for M1-$VO_2$ (Fig. 1f), consistent with the experimentally observed value [4,59].

The atomic configuration was equilibrated for 1 ps using a time step of 1 fs at 300 K in the NVE ensemble during BOMD simulations. Selecting the thermally equilibrated structure at 300 K as the initial configuration for TDDFT calculations. To simulate photoinduced IMT dynamics, the time-space external field can be described as a Gaussian shape,

$$E(t) = E_0 \cos(\omega t) \exp[-(t-t_0)^2/(2\sigma^2)]$$

where we selected photon energy of $\omega$ = 1.55 eV, consistent with experimental parameters [4,28], a time delay of $t_0$ = 25 fs, and a width parameter $\sqrt{2}\sigma$ = 10 fs. The electric field $E_0$ is represented by the A field, and precisely tuned to achieve varying electronic excitations at 300 K, as illustrated in Supplementary Figs. 2a, b.

## Data availability

The data that support the findings of this study are available from the corresponding authors upon reasonable request.

## Code availability

The rt-TDDFT CODE has been integrated into the PWmat package. The PWmat software can also be accessed directly from http://www.pwmat.com.

**ACKNOWLEDGMENTS**

The work was supported by the National Natural Science Foundation of China (NSFC) under Grant Nos. 11925407, 61927901, and 12174380, the Key Research Program of Frontier Sciences, CAS under Grant No. ZDBS-LY-JSC019, CAS Project for Young Scientists in Basic Research under Grant No. YSBR-026, the Strategic Priority Research Program of the Chinese Academy of Sciences under Grant No. XDB43020000, the key research program of the Chinese Academy of Sciences under Grant No. ZDBS-SSW-WHC002, and the China Postdoctoral Science Foundation under Grant No. 2022M723073.

**Author Contributions**

F.G. performed the TDDFT simulations and prepared the figures with the help of W.L. F.G. and W.L. conducted the analysis, discussion, and wrote the manuscript. J.L. and L.W. proposed the research project, established the project direction, and revised the paper with input from F.G. and W.L. All authors contributed to the analysis and discussion of the data. F.G. and W.L. contributed equally to this work.

**Competing interests**

The authors declare no competing interests.

Supplemental Materials for

# The photoinduced hidden metallic phase of monoclinic VO$_2$ driven by local nucleation via a self-amplification process

Feng-Wu Guo[1,2]†, Wen-Hao Liu[1]†, Zhi Wang[1], Shu-Shen Li[1,2], Lin-Wang Wang[1]*, and Jun-Wei Luo[1,2]*

†These authors contributed equally to this work.
*Email: lwwang@semi.ac.cn; jwluo@semi.ac.cn

**Note 1: RT-TDDFT simulations**

**Note 2: Evolution of Bragg peaks**

**Note 3. Calculation of Transmission Spectra**

**Note 4. Photoexcited Carrier Distributions and Atomic Driving Force on V-V dimers**

**Note 1: RT-TDDFT simulations**

The real-time time-dependent density functional theory (rt-TDDFT) calculations in this study employed the local density approximation functional (LDA) + U exchange-correlation with a Hubbard U value of 3.4 eV. The simulation model consisted of a 96-atom supercell of the M1-phase, utilizing a plane-wave basis with a cutoff energy of 45 Ry. To ensure accurate Brillouin-zone sampling, a 2×2×2 Monkhorst-Pack k-point mesh was employed.

The atomic configuration was equilibrated for 1 ps using a time step of 1 fs at 300 K in the NVE ensemble during Born-Oppenheimer Molecular Dynamics (BOMD) simulations. The thermally equilibrated structure at 300 K was selected as the initial configuration for the time-dependent density functional theory (TDDFT) calculations. Within this rt-TDDFT algorithm [1], the time-dependent wave functions, $\varphi_i(t)$, were expanded by the adiabatic eigenstates, $\phi_l(t)$.

$$\varphi_i(t) = \sum_l C_{i,l}(t)\, \phi_l(t) \quad (1)$$

and

$$H(t)\phi_l(t) \equiv \varepsilon_l(t)\phi_l(t) \quad (2)$$

In this context, $H(t) \equiv H(t, R(t), \rho(t))$, where $R(t)$ denotes the nuclear positions, and $\rho(t)$ represents the charge density. Employing equation (1), the evolution of the wave functions $\varphi_i(t)$ is reformulated into the evolution of the coefficients $C_{i,l}(t)$. In equation (2), a linear time-dependent Hamiltonian is implemented to capture the time evolution of the Hamiltonian within each time step. Consequently, our rt-TDDFT simulation utilizes a significantly larger time step (0.1 fs).

To emulate photoexcitation, a uniform A field in reciprocal space is introduced [2].

$$H = 1/2(-i\nabla + A)^2 = 1/2(-i\nabla_x + A_x)^2 + 1/2(-i\nabla_y + A_y)^2 + 1/2(-i\nabla_z + A_z)^2 \quad (3)$$

The time-space external field can be described as a Gaussian shape,

$$E(t) = E_0 \cos(\omega t)\, exp[-(t - t_0)^2/(2\sigma^2)] \quad (4)$$

where we selecte a photon energy of $\omega$ = 1.55 eV, a time delay of $t_0$ = 25 fs, and a width parameter $\sqrt{2}\sigma$ = 10 fs. The electric field is adjusted to $E_0$ = 0.13, 0.27, 0.40, 0.45, and 0.53 V/Å at 300 K to attain electronic excitations of 0.11%, 0.39%, 0.79%, 0.97% and 1.25% in free structure. When frozen structure, the laser fluence is set to $E_0$ = 0.13, 0.27, 0.40, 0.45, 0.53, 0.80, 1.07 and 1.31 V/Å to achieve electronic excitations of 0.11%, 0.39%, 0.79%, 0.97%, 1.25%, 2.10%, 3.00% and 3.75%, as illustrated in Fig. S2a-b.

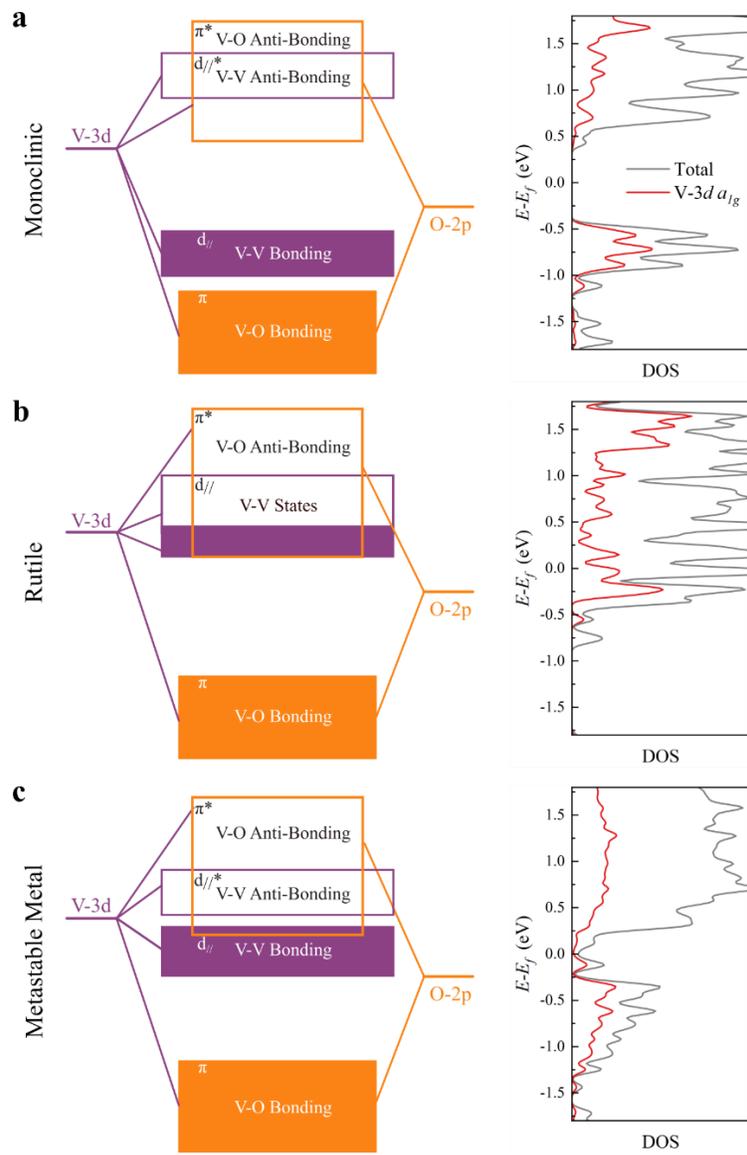

Fig. S1. Schematic of the bonding and antibonding states and the total density of states (DOS) of the M1 phase (a), the R phase (b) and the MM phase (c).

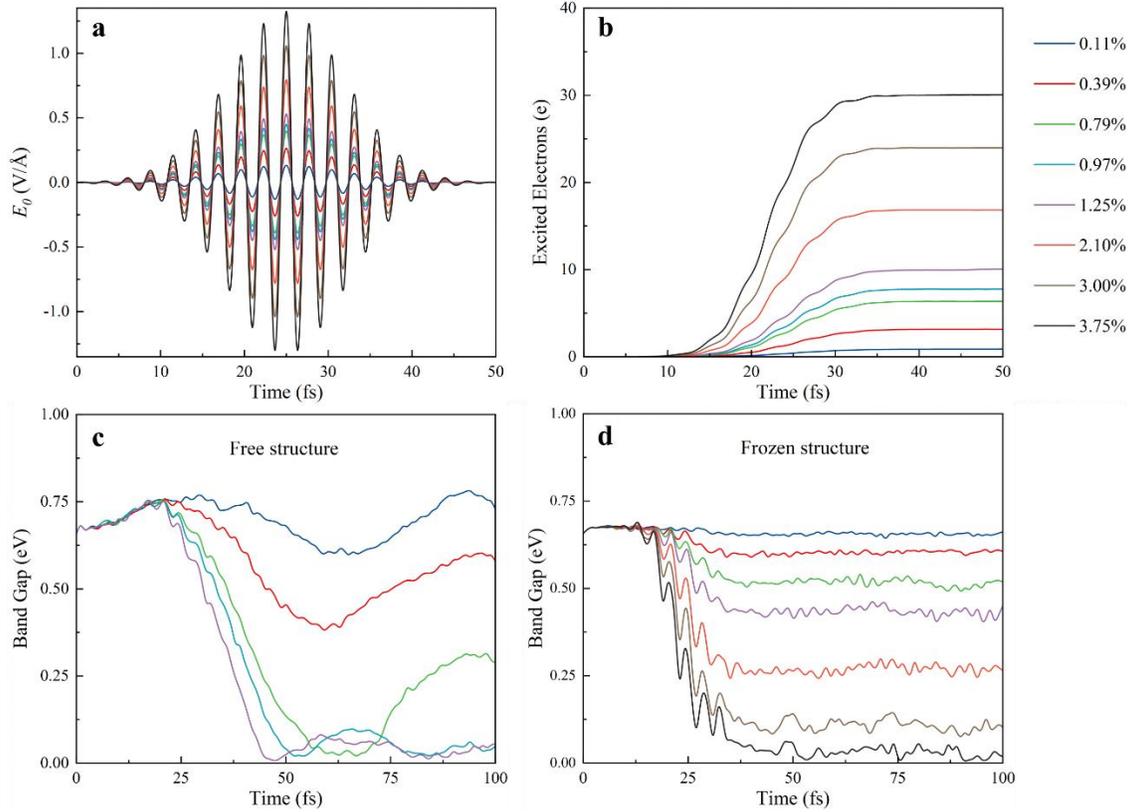

Fig. S2. (a) The profile of the external electric field applied to $VO_2$. The photon energy is 1.55 eV [3,4]. (b) The varying number of photoexcited electrons transitioning from the valence bands to conduction bands under the laser pulses described in (a). The dynamic evolution of the band gap is depicted for the free structure (c) and the frozen structure (d) under different photoexcitation.

**Note 2. Evolution of Bragg peaks**

The simulated X-Ray diffraction spectrum (XRD) was generated using the crystallographic simulation software VESTA based on the structures in our dynamic simulations. By comparing with experimental data [5], we initially determined the positions of the 200 and 220 peaks. Given the low intensity of the $30\bar{2}$ peak or other peaks characterizing the transition from the M1 phase to the R phase, we determined the $30\bar{2}$ peak in ultrafast electron diffraction experiments using the equation:

$$q = 4\pi \sin\theta / n\lambda$$

In our simulation, $\lambda = 1.54059$ Å. Subsequently, we can determine $q$ in ultrafast electron diffraction by establishing the relationship between $\theta$ and $q$. Through correlation, we determine XRD positions: 38.55° (200 peak, corresponding to 0.21 Å$^{-1}$ in UED), 52.3° ($30\bar{2}$ peak, corresponding to 0.27 Å$^{-1}$ in UED), and 57.3° (220 peak, corresponding to 0.29 Å$^{-1}$ in UED), aligning with experimental data [3]. Here, given the presence of three similar peaks around 52.3° in our simulated XRD (Fig. S3), and their consistent changes during the structural phase transition (Fig. S4), we designate the first peak as the $30\bar{2}$ peak.

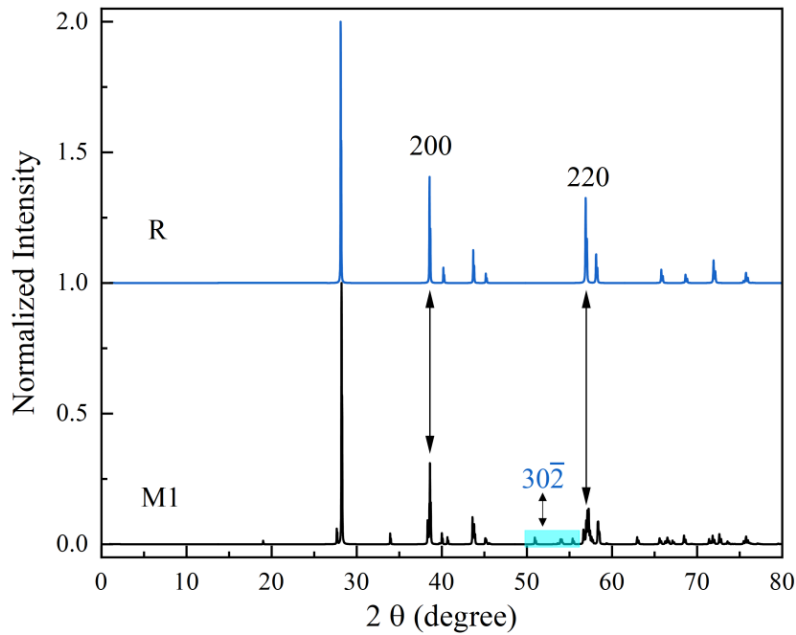

Fig. S3. X-Ray diffraction spectrum (XRD). The blue line corresponds to the XRD pattern of the ideal lattice in the R phase, and the black line represents the XRD pattern of the ideal lattice in the M1 phase. The positions of the 200, 200, and $30\bar{2}$ peaks align with experimental observations [3,5]. The disappearance of the $30\bar{2}$ peak and the increase of the peaks at 200 and 220 mark the phase transition from M1-to-R phase.

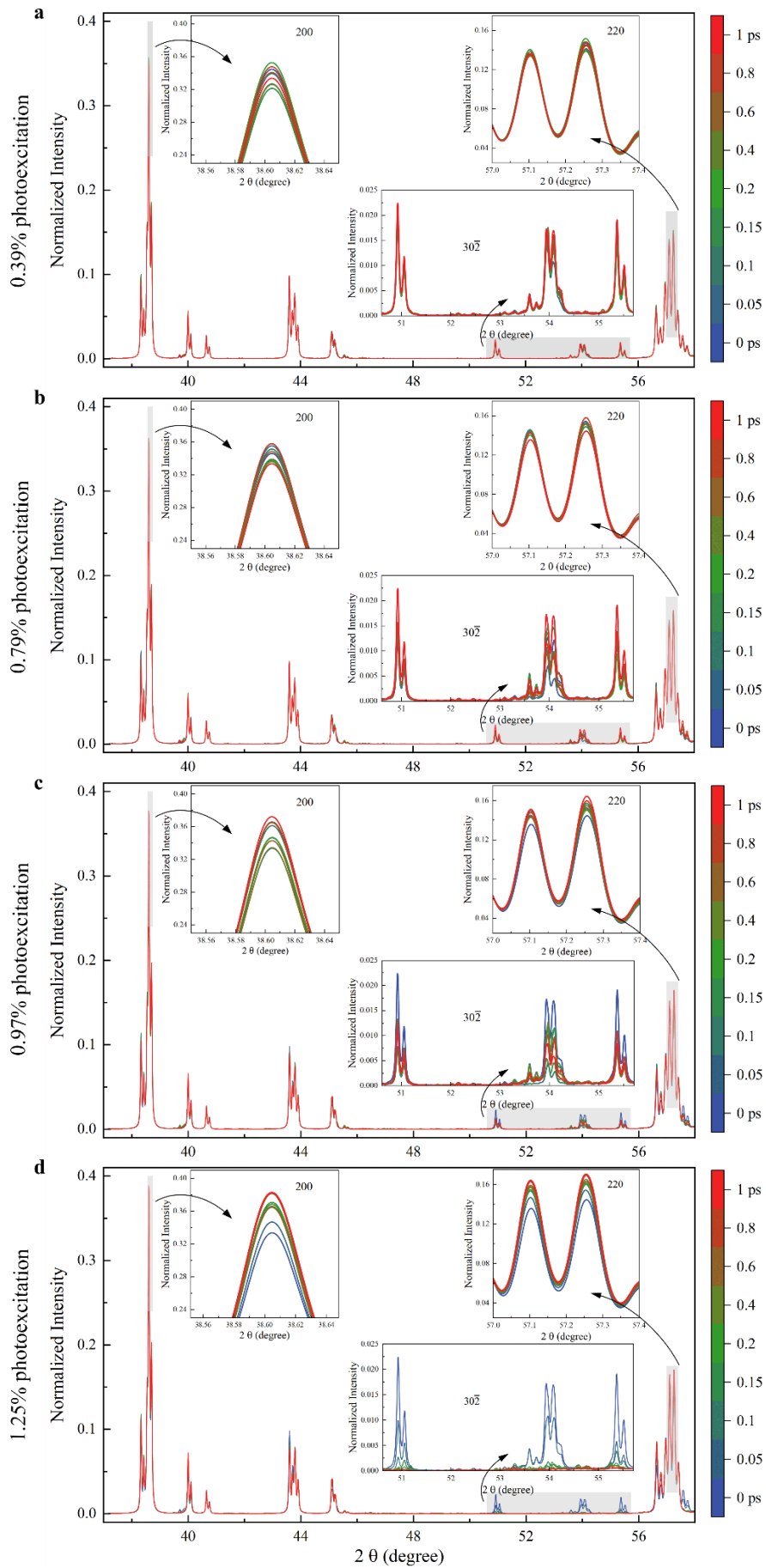

Fig. S4. Dynamic evolution of XRD under electronic excitations of 0.39% (a), 0.79% (b), 0.97% (c), and 1.25% (d). Insets provide partial enlargements of the 200, 220, and 30$\bar{2}$ peaks. In all cases, the peaks at 220 and 200 exhibit an increase. The 30$\bar{2}$ peak completely disappears only when the excited electron concentration reaches 1.25%. All structures mentioned here are derived from rt-TDDFT simulations. The XRD patterns at a given moment are obtained from the crystal structure at the corresponding time.

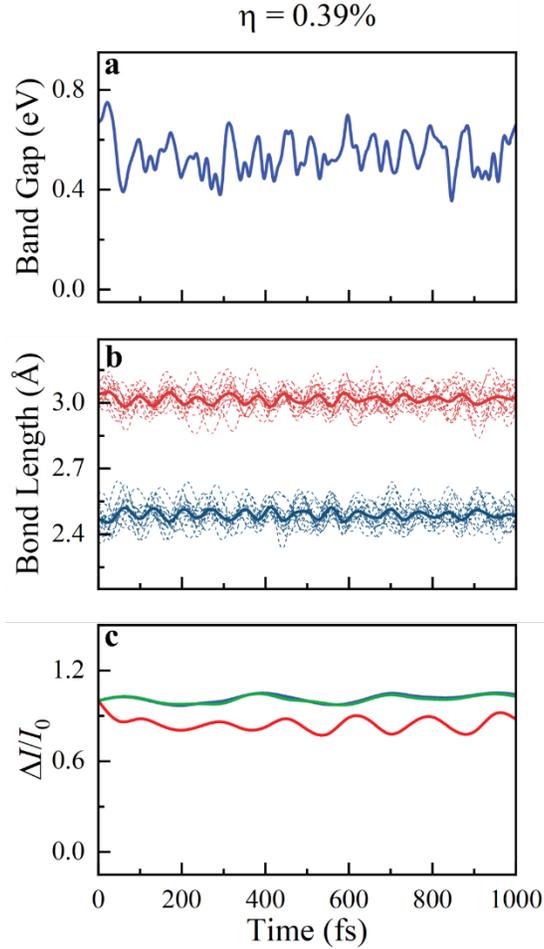

Fig. S5. The dynamic evolution of the band gap (a), V-V bond lengths (b) and Bragg peaks (c) at electronic excitations of η = 0.39%. η represents the proportion of excited electrons to the total number of electrons. The red and blue dashed lines in (b) correspond to the long and short bond lengths, respectively. The average bond lengths of long and short bonds are represented by red and blue solid lines. Evolution of the simulated 200 (blue lines), 220 (green lines), 30$\bar{2}$ (red lines) Bragg peaks (c). The band gap experiences a slight decrease and the characteristic of all V-V long and short bonds in M1 phase are almost maintained. The 30$\bar{2}$ peak only drops by about 15% without the emergence of a phase transition [6].

**Note 3: Calculation of Transmission Spectra**

To more accurately elucidate the characteristics of excited-state electrons, we extended the Random Phase Approximation (RPA) method, originally tailored for computing absorption spectra in the ground state. The imaginary component of the frequency-dependent dielectric function is delineated as follows:

$$\epsilon_\infty^{(2)}(\omega) = 2\frac{4\pi^2}{\Omega}\frac{1}{\omega^2}\frac{1}{N_k}\sum_{n,m,k}(f_{nk} - f_{mk})|\lambda \cdot p_{nm}|^2 \delta(\varepsilon_{mk} - \varepsilon_{nk} - \hbar\omega) \quad (5)$$

Here, in the ground state, $f_{nk}$ is defined by Fermi-Dirac distribution at 300 K. The matrix $p_{nm} = \langle n|\vec{p}|m\rangle$, where $|n\rangle$ and $|m\rangle$ are wave functions obtained through self-consistent calculations. Notably, when $f_{nk}$ denotes the occupation of excited-state electrons at the *k*-th K point in the *n*-th energy band, and the $|n\rangle$ and $|m\rangle$ in $p_{nm} = \langle n|\vec{p}|m\rangle$ represent the wave functions of the excited states corresponding to the *n*-th and *m*-th energy bands, it is suitable for describing the excited state. Both the occupation of electrons and the wave function at a certain moment in the corresponding dynamical process are generated through rt-TDDFT calculations. This comprehensive approach ensures an accurate representation of the excited-state dynamics in the system. The real part is obtained by Kramers-Kronig transformation from $\epsilon_\infty^{(2)}(\omega)$:

$$\epsilon_\infty^{(1)}(\omega) = 1 + \frac{1}{\pi}\int_0^\infty \frac{\epsilon_\infty^{(2)}(\omega')\omega'}{\omega'^2 - \omega^2}d\omega' \quad (6)$$

The extinction coefficient is:

$$\kappa(\omega) = [\frac{\sqrt{\epsilon_1^2 + \epsilon_2^2} - \epsilon_1}{2}]^{\frac{1}{2}} \quad (7)$$

Finally, the absorption coefficient is obtained as

$$\alpha(\omega) = \frac{4\pi}{\omega}\kappa(\omega) \quad (8)$$

The simulated absorption coefficients are illustrated in Fig. S6 and Fig. S7. To derive the infrared transmission spectrum at 0.25 eV, we extract the absorption coefficients at 0.25 eV for different time. The transmission spectrum, denoted as *T*, is computed using the formula:

$$T = 1 - \text{normalized}(\alpha_t(0.25)) \quad (9)$$

where the absorption coefficients are normalized at 0.25 eV.

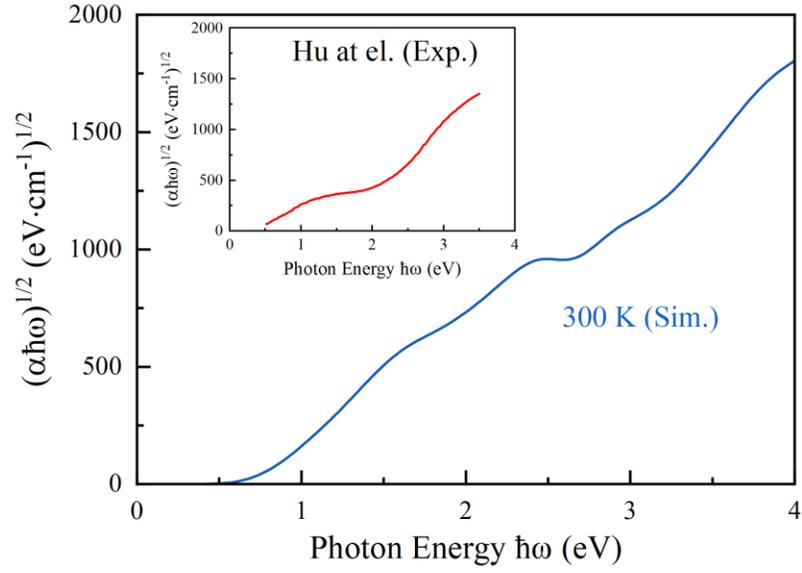

Fig. S6. The absorption spectra of the M1 phase at 300 K in the ground state. The experimental data are shown in the inset. Our simulated result of RPA method is in agreement with the experimental observations [7].

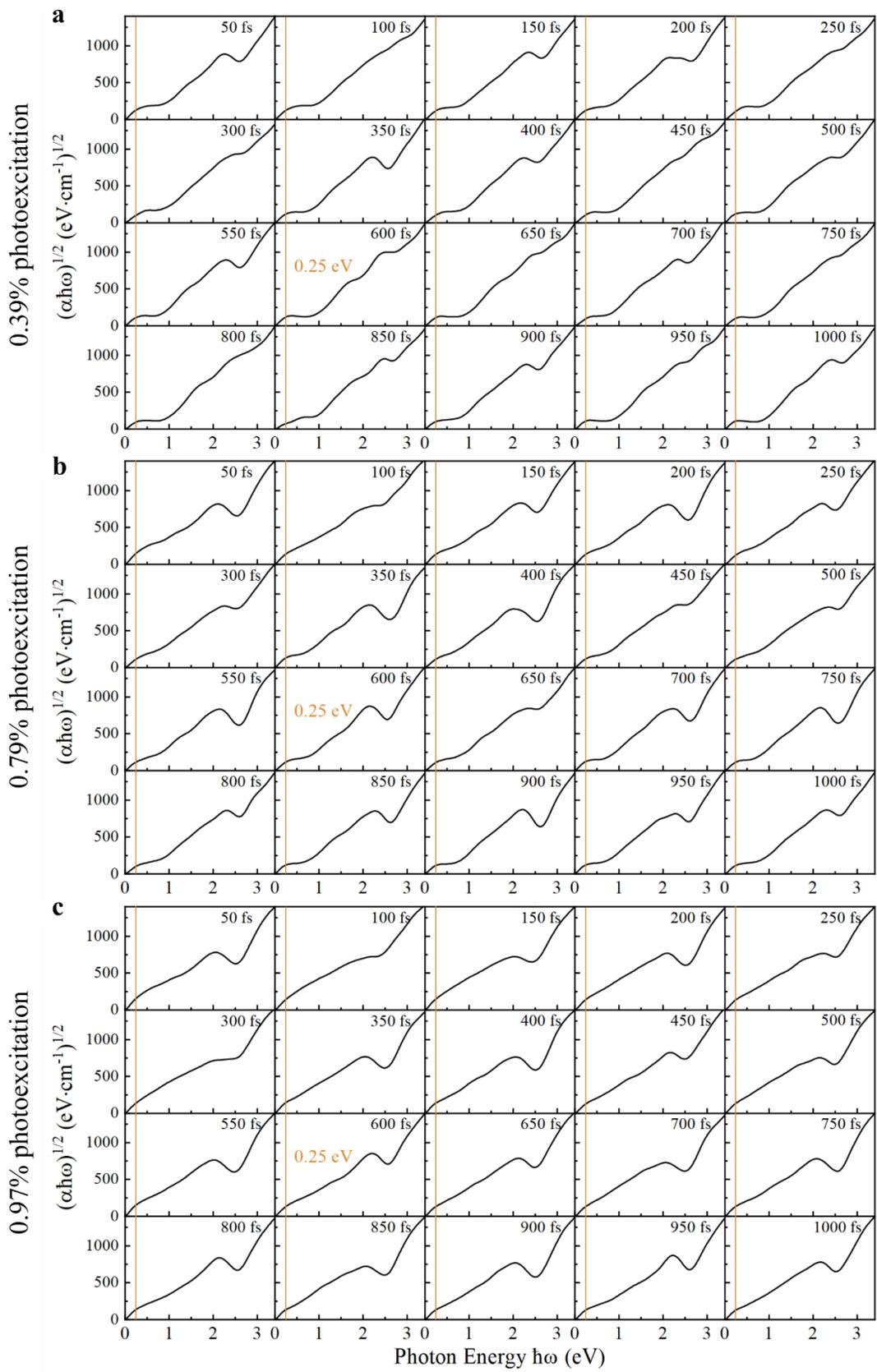

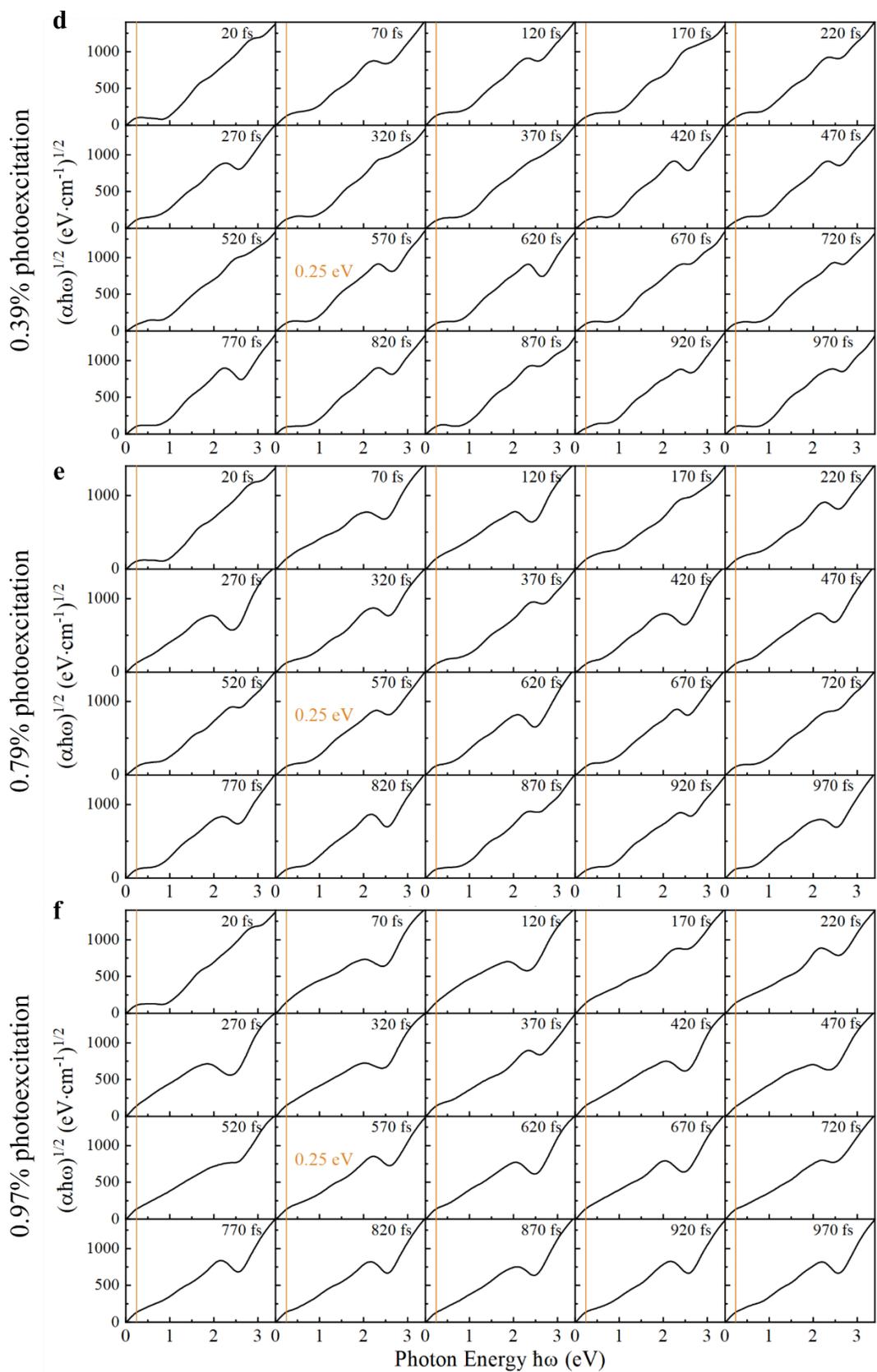

Fig. S7. Temporal evolution of absorption spectra under electronic excitations of 0.39% (a,d), 0.79% (b,e),

and 0.97% (c,f). The orange lines depict the data at 0.25 eV, crucial for calculating the transmission spectrum in Fig. S8.

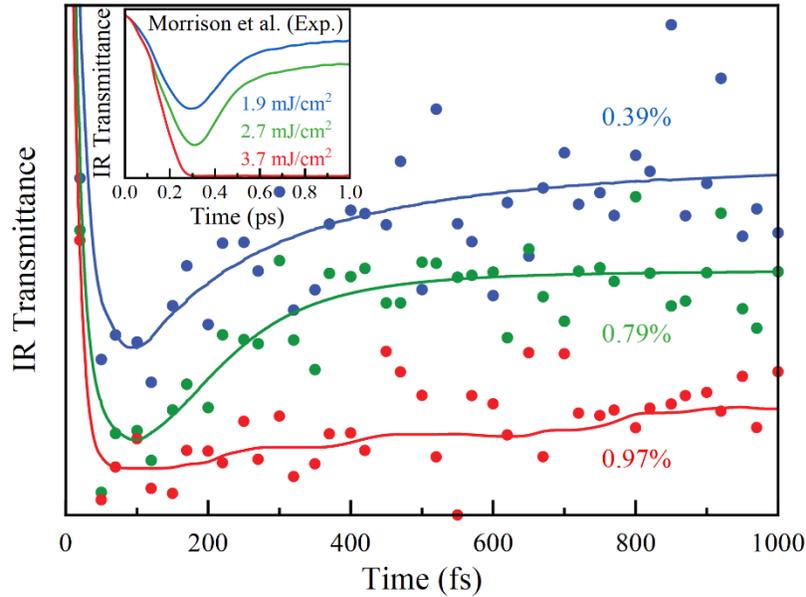

Fig. S8. Time-resolved IR transmissivity simulations. Time-resolved IR (5 mm, 0.25 eV) transmissivity at electronic excitations of $\eta = 0.39\%$, $\eta = 0.79\%$, and $\eta = 0.97\%$, depicted by blue, green, and red circles, respectively.

Our calculations of absorption spectra of the photoexcited states at different times are presented in Fig. S7. We then extract the values of the absorption spectra at photon energy $\hbar\omega$ = 0.25 eV [3] to generate the transmission spectrum (Fig. S8). At electronic excitation of 0.39% and 0.79%, the IR transmission exhibits a trend of first reduction and then rapid recovery, consistent with experimental observations [3]. The change in IR transmission can be explained by the evolution of band structures, in which more bands cross to the Fermi level at 60 fs (Figs. S9a, b), resulting in a larger photon absorption and consequently a minimum IR transmission. Subsequently, the bands near the Fermi level gradually sparse, leading to a reduction in absorption and an enhancement in IR transmission. Specifically, under 0.79% excitation, there is a more significant deviation of the bands near the Fermi level compared to the time point at 60 fs, resulting in a more pronounced recovery of transmission compared to 0.39% excitation. This observation further elucidates why a greater recovery is experimentally observed at a higher laser fluence of 2.7 mJ/cm² compared to 1.9 mJ/cm² [3]. At an electronic excitation of 0.97%, the IR transmission reaches a minimum and almost remains at that value due to the long-lived existence of metallic properties with an isolated band near the Fermi level (Fig. S9c), which also aligns well with experimental results [3].

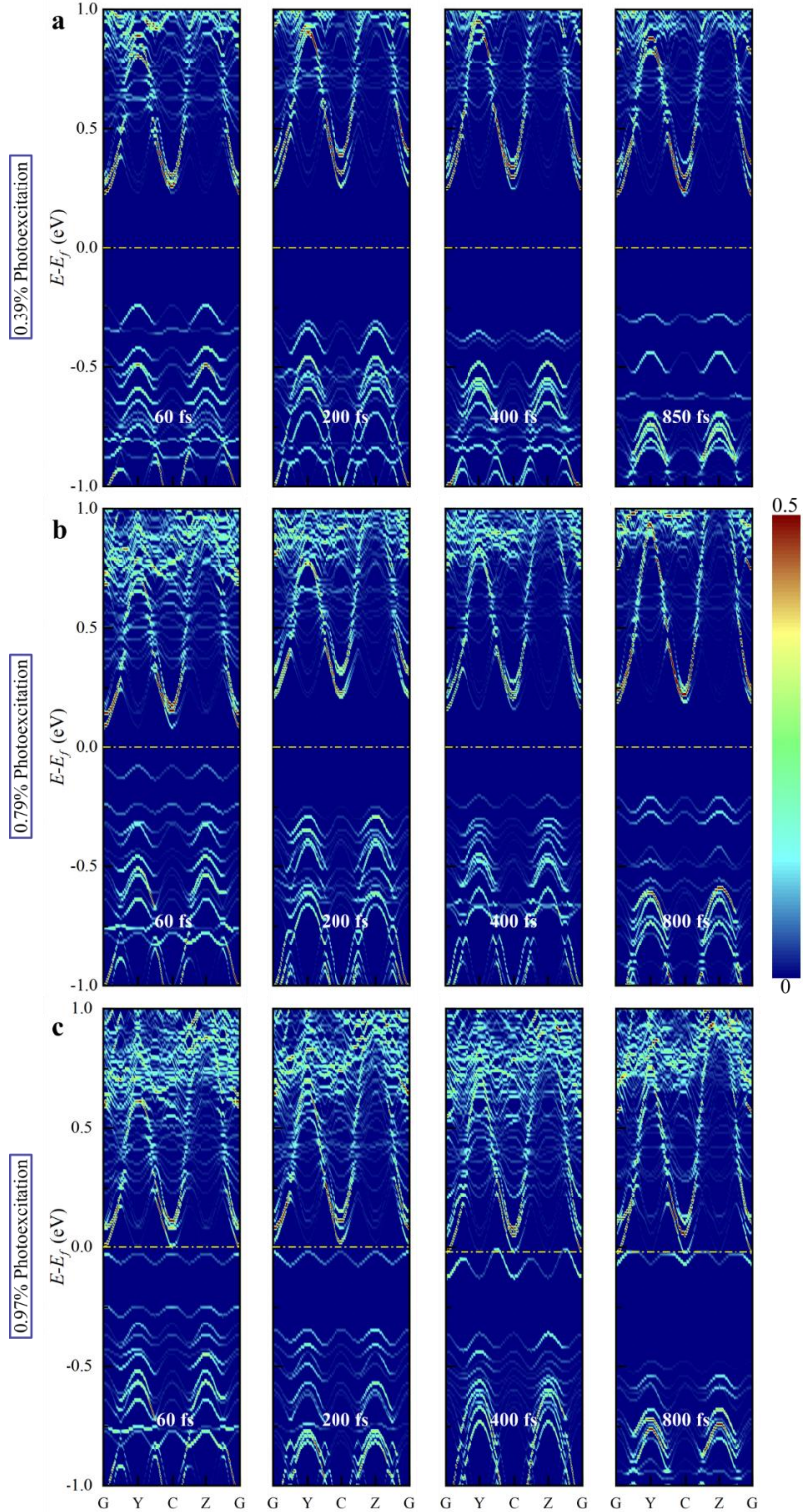

Fig. S9. Temporal Evolution of Unfolded Band Structures. Calculated unfolded band structures at electronic excitations of 0.39 (a), 0.79% (b) and 0.97% (c). The isolated energy level around the Fermi level rapidly emerges and disappears at electronic excitation of 0.79%, but remains stable at an electronic excitation of 0.97% throughout the entire dynamics.

**Note 4. Photoexcited Carrier Distributions and Atomic Driving Force on V-V dimers**

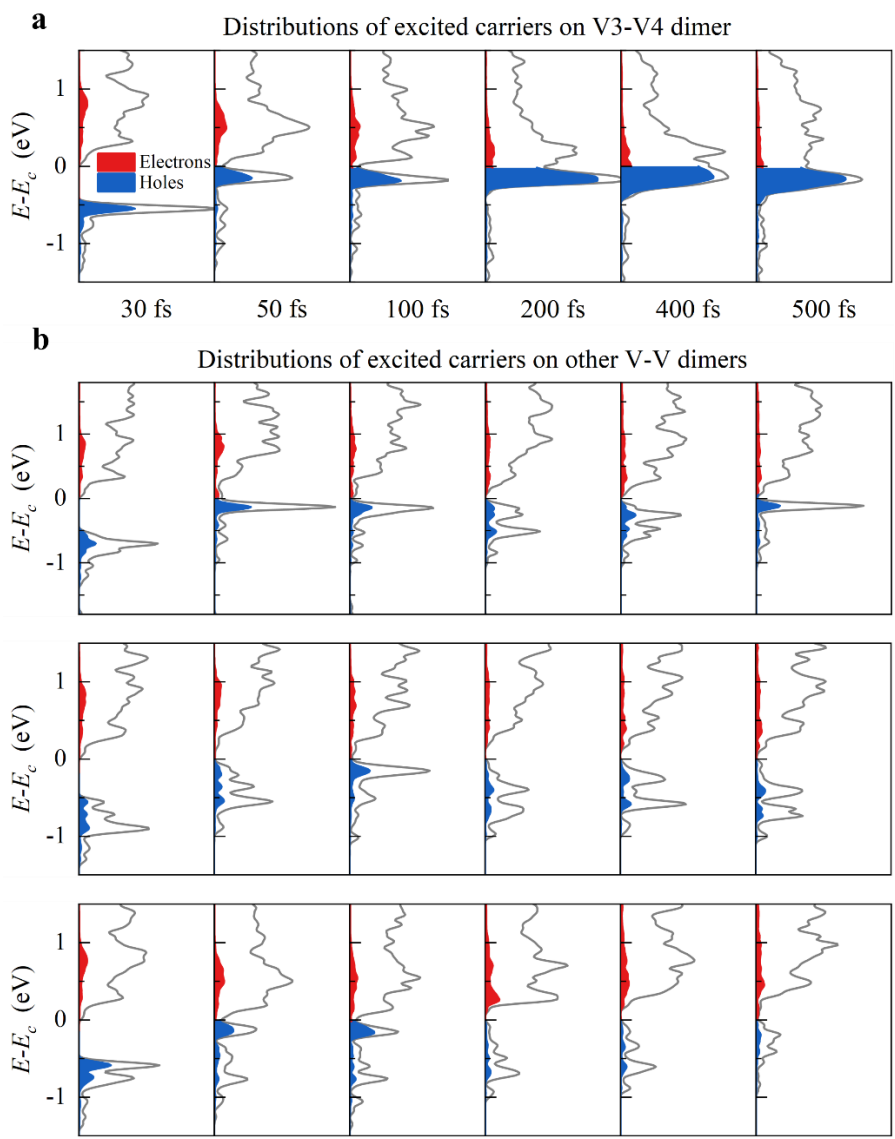

Fig. S10. Evolution of excited carriers of V3-V4 dimer (a) and other V-V dimers (b) at 0.97% elelctronic excitations. Red (blue) shaded areas represent photoexcited electrons (holes).

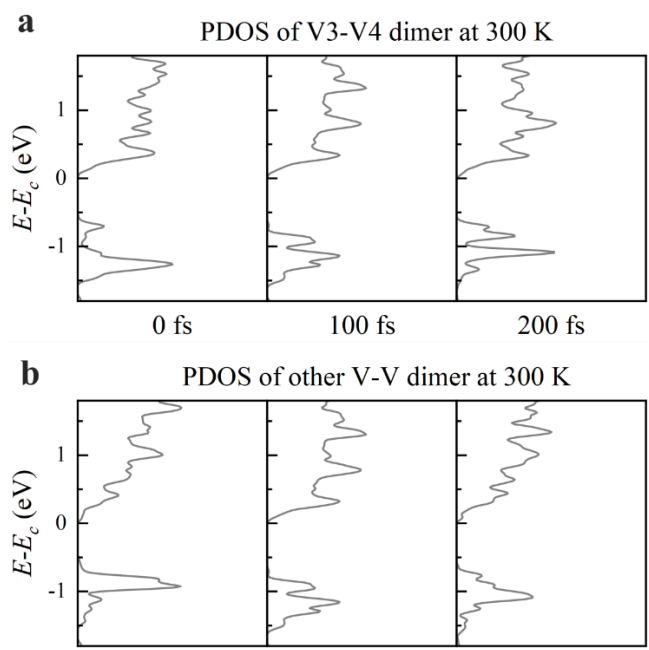

Fig. S11. The evolution of the V3-V4 dimer (a) and other V-V dimers (b) under no electronic excitations at 300 K. The bonding states of V-V dimers can be broadened due to atomic thermal vibrations, but this does not affect the characteristic of V-V dimers.

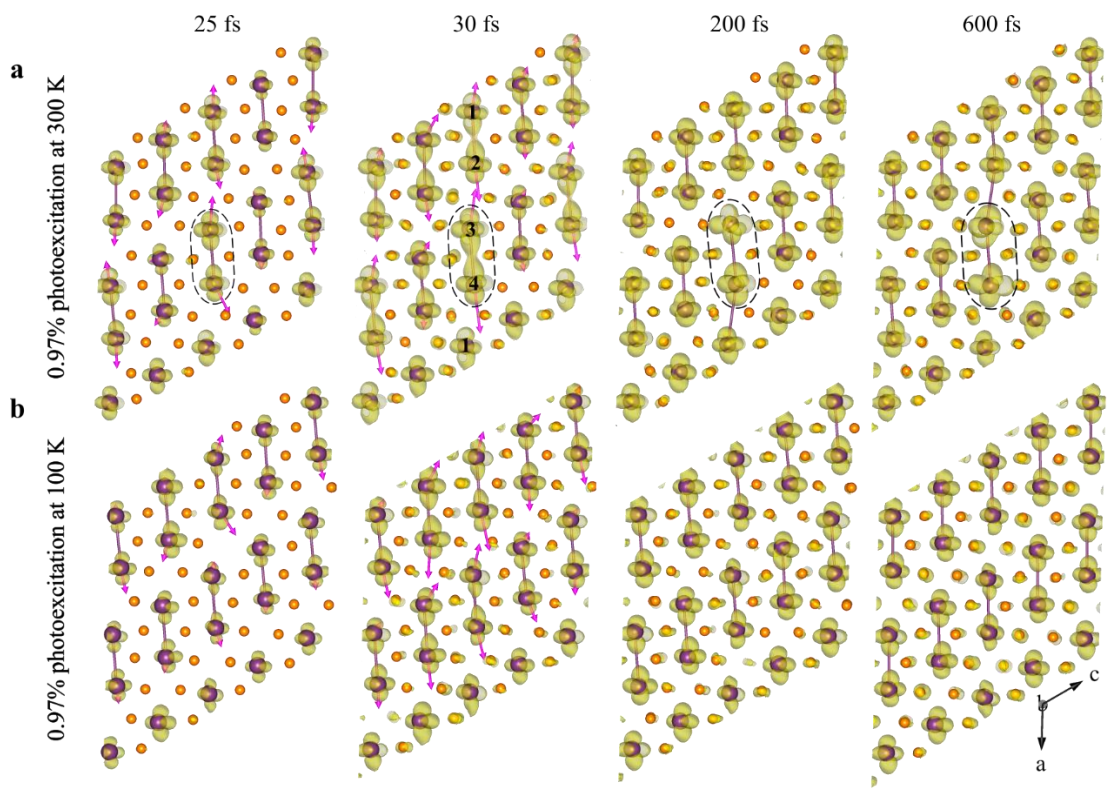

Fig. S12. The dynamical evolution of photoexcited hole density and force distribution on V atoms for electronic excitations of 0.97% at 300 K (a) and 0.97% at 100 K (b). The shaded area represents the distribution of photoexcitation holes, while the pink arrows indicate the photoexcitation force. The black dashed ellipse (a) indicates the local structural phase transition.

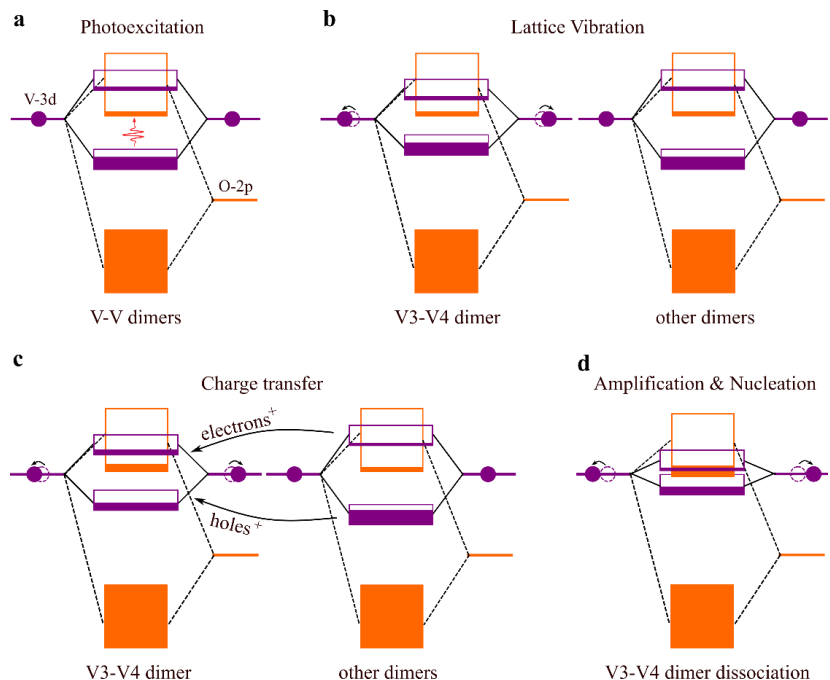

Fig. S13. The schematic diagram of the self-trapping and self- amplified dynamics. **a**, Initially, the incident photon triggers the transition of electrons from the V-V bonding state to both the V-V anti-bonding state and the V-O anti-bonding state. **b**, Over time, V-V dimers experience small fluctuations induced by thermal vibrations, then local V3-V4 dimer with slightly longer lengths will lower (raise) the V-V antibonding (bonding) state more than other dimers. **c**, The V3-V4 dimer will trap more photoexcited holes from other dimers due to its higher bonding state, causing a larger driving force on the V3-V4 dimer to futher elongate the bond length. **d**, This elongation, in turn, causes the rise of the energy level of bonding states and the reduction of the energy level of antibonding states. This induces self-trapping and self-amplifying dynamics as the excited holes will relax to the highest valence state, and further accumulation of the holes on the V3-V4 dimer will further raise its bonding state level. Finally, the V3-V4 dimer with more photoexcited holes will be dissociated to form local phase transition.

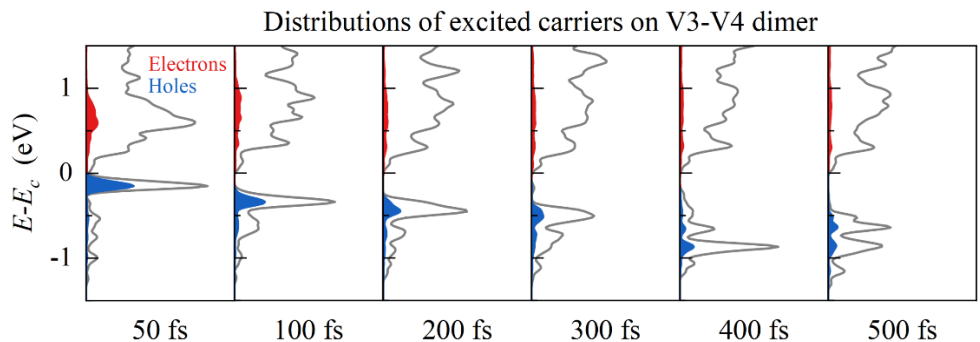

Fig. S14. PDOS of V3-V4 dimer from rt-TDDFT simulation at different time at 0.79% electronic excitations.